\documentclass{article}
\usepackage{spconf,amsmath,graphicx,hyperref, multirow, bm, subcaption, makecell, float}
\usepackage[table]{xcolor}
\usepackage{adjustbox}

\title{Neural Acoustic Multipole Splatting \\for Room Impulse Response Synthesis}
\name{Geonwoo Baek$^1$ and Jung-Woo Choi$^1$%
\sthanks{Corresponding author.}}
\address{
$^1$School of Electrical Engineering, KAIST, Daejeon, Republic of Korea \\
\{bkw6287, jwoo\}@kaist.ac.kr
}

\begin{document}
\ninept
\maketitle

\begin{abstract}
Room Impulse Response (RIR) prediction at arbitrary receiver positions is essential for practical applications such as spatial audio rendering. We propose Neural Acoustic Multipole Splatting (NAMS), which synthesizes RIRs at unseen receiver positions by learning the positions of neural acoustic multipoles and predicting their emitted signals and directivities using a neural network. Representing sound fields through a combination of multipoles offers sufficient flexibility to express complex acoustic scenes while adhering to physical constraints such as the Helmholtz equation.
We also introduce a pruning strategy that starts from a dense splatting of neural acoustic multipoles and progressively eliminates redundant ones during training. Experiments conducted on both real and synthetic datasets indicate that the proposed method surpasses previous approaches on most metrics while maintaining rapid inference. Ablation studies reveal that multipole splatting with pruning achieves better performance than the monopole model with just 20\% of the poles.


\end{abstract}
\section{Introduction}
\label{sec:intro}
Room Impulse Responses (RIRs) characterize how sound propagates in a room, including direct sound, early reflections, and late reverberation. Accurate RIR rendering is crucial for spatial audio, virtual and augmented reality, and interactive gaming, as it provides realistic and immersive room cues at various listener positions. In reality, we can only obtain a limited number of RIRs, but we need to simulate them for numerous unseen receiver locations. This paper tackles the fundamental challenge of predicting RIRs at unseen positions using a limited dataset of measured or simulated RIRs.

RIR estimation has been addressed through multiple approaches. Notable examples include modal expansion~\cite{haneda1999common, das2021room}, the equivalent source method (ESM)~\cite{tsunokuni2021spatial, matsuhashi2023spatial, antonello2017room}, and plane wave expansion (PWE)~\cite{antonello2017room, hahmann2022convolutional, jin2015theory}-based methods, all aimed at solving the acoustic inverse problem using regularizations and constraints. These traditional techniques explicitly model physical structures to recreate sound fields. Yet, accuracy typically suffers at high frequencies, and practical application to wideband audio is hampered by ill-conditioning and aliasing artifacts. Additionally, RIR parameterization techniques~\cite{raghuvanshi2018parametric, raghuvanshi2014parametric} have been introduced to model RIRs using several room acoustic parameters, which enable dynamic acoustic effects in gaming. However, these estimations can be suboptimal in addressing all aspects of RIRs.

Recently, deep neural networks (DNNs) have gained traction for this task. Initially, DNNs estimated RIRs from given source and receiver positions \cite{luo2022learning, su2022inras} but lacked physical consistency and struggled with high-frequency components. Other approaches introduced physical constraints to enhance model stability and generalization~\cite{karakonstantis2024room, masuyama2025physics, pezzoli2023implicit, kurata2024noise}. They integrated physical priors, like the wave equation, into the loss function to guide learning~\cite{karakonstantis2024room, masuyama2025physics}. For better early reflection predictions, some models used periodic activation functions~\cite{pezzoli2023implicit, sitzmann2020implicit}, while others applied the dynamic pulling method to enhance performance in noisy settings~\cite{kurata2024noise}. These models need only a few RIRs for training~\cite{karakonstantis2024room} and ensure physical fidelity, yet training can be slow due to PDE residual computation, may face gradient issues, and exhibit limited high-frequency accuracy~\cite{bi2024point}.

Recently, approaches that incorporate pre-designed physical models into the network optimization have been suggested~\cite{lan2024acoustic, bi2024point}. These approaches develop a physical model inherently satisfying necessary physical constraints and estimate model inputs or parameters that align with the training data.
Such approaches offer improvements over previous methods, particularly in capturing high-frequency components~\cite{bi2024point}. 
For example, AVR~\cite{lan2024acoustic} models the sound field as the combination of signals emitted from points on rays surrounding the receiver and considers the attenuation and time delay caused by sound propagation through rays. The point neuron model~\cite{bi2024point} uses point neurons satisfying the Helmholtz equation as basis functions to describe the sound field and trains their signals and positions to encode and estimate the entire field. 
These model-fitting approaches aligning the basis functions to the observed signal are in line with Gaussian Splatting (GS)~\cite{kerbl20233d} that utilizes Gaussian functions for 3-D field reconstruction from 2-D vision images.

Existing acoustic model-fitting methods often suffer from inefficiencies of basis functions like points on rays or acoustic monopoles, which require a large number to function effectively, thereby increasing inference time. For instance, the image source method~\cite{allen1979image} can effectively mimic early reflections from specular surfaces using monopoles but struggles with simulating scattering from walls and objects. Such irregular reflections can be better represented as directional multipoles. While compact monopoles can also act as directional multipoles, achieving high-order directivities leads to amplified monopole signals, causing instability in DNN optimizations. 

To address this, we introduce a multipole-based RIR estimation network, NAMS, which uses acoustic multipoles with adjustable directivities to model and estimate RIRs. This model spatially distributes multipoles, optimizes their positions, and predicts both their signals and frequency-dependent directivities. In addition, inspired by adaptive density control of GS~\cite{kerbl20233d}, we propose a pruning strategy that optimizes the number of multipoles during training. Initially, we densely splat multipoles and progressively prune redundant ones throughout training. Our experimental results conducted in real and virtual environments reveal that NAMS outperforms existing methods in terms of estimated room acoustic parameters while achieving faster inference speeds.

\begin{figure*}[t]
    \centering
    \includegraphics[width=0.95\linewidth]{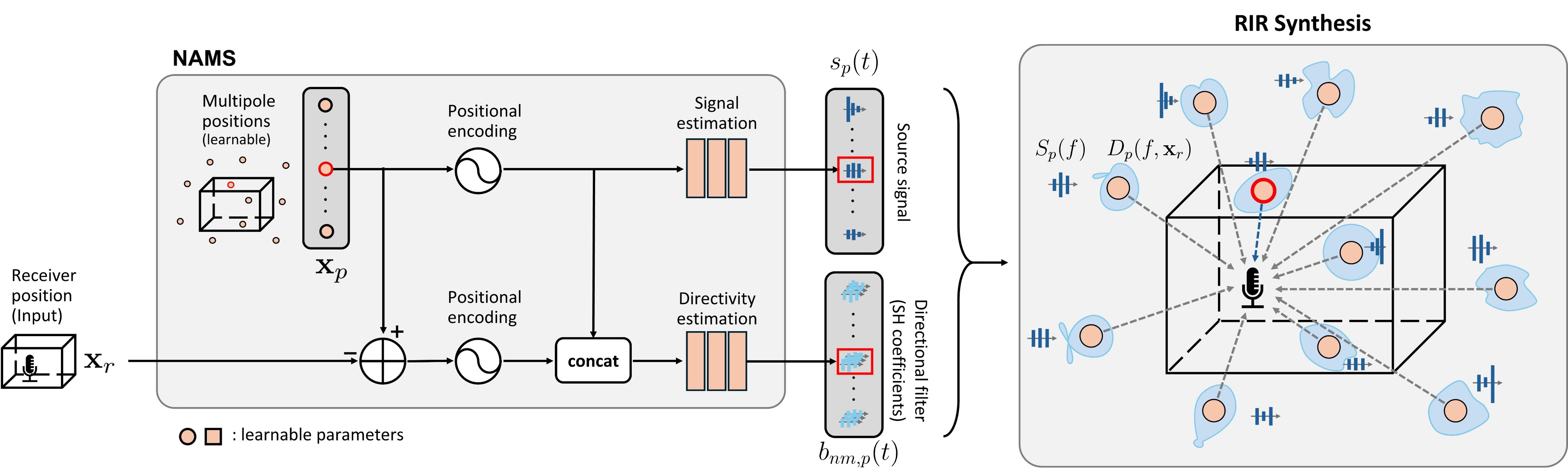}
    \caption{\textbf{Overview of NAMS framework.} We optimize each neural acoustic multipole position $\mathbf{x}_p$ and train MLPs to predict its emitted signal $s_p(t)$ and spherical harmonic coefficients $b_{nm,p}(t)$. We synthesize the RIR using the receiver position $\mathbf{x}_r$ and the set of neural acoustic multipole positions, emitted signals, and spherical harmonic coefficients.}
    \label{fig:placeholder}
    \vspace{-10pt}
\end{figure*}

\section{PROPOSED METHOD}
\label{sec:method}

\subsection{NAMS for Room Impulse Response Synthesis}
\label{sec: neural multipoles}
NAMS generates RIRs by employing multipoles that have adjustable directional patterns and emit signals of limited duration. The directional patterns $D_p(f,\mathbf{x}_r)$ for the $p$-th multipole at frequency $f$ can be expressed in terms of the spherical harmonics~\cite{gumerov2005fast} as  
\begin{eqnarray}
D_p(f,\mathbf{x}_r)&=&\sum_{n=0}^N \sum_{m=-n}^n B_{nm,p}(f)\, Y_n^m(\mathbf{\Omega}_{p}(\mathbf{x}_r)),\label{eq:dir_f}
\end{eqnarray}
where $n\le N$ and $m$ denote the order and degree of spherical harmonics $Y_n^m$, and $\mathbf{\Omega}_p(\mathbf{x}_r)$ indicates the angular position of the multipole at $\mathbf{x}_p$ measured from the receiver position $\mathbf{x}_r$. The coefficients $B_{nm,p}$ determine the directional pattern of the $p$-th multipole. 

We model the RIR as the superposition of these multipoles, each of which emits a signal $S_p(f)$. With a far-field assumption, the RIR in frequency domain can be approximated as  
\begin{eqnarray}
H(f,\mathbf{x}_r) &=& \sum_{p=1}^P S_p(f)
\frac{e^{-j 2 \pi f r_p(\mathbf{x}_r) / c}}{r_p(\mathbf{x}_r)}\;
D_p(f,\mathbf{x}_r), \label{eq:model_f} 
\end{eqnarray}
where ${e^{-j 2 \pi f r_p / c}}$ expresses the propagation delay for the speed of sound $c$, and $1/r_p$ represents attenuation across the propagation distance $r_p(\mathbf{x}_r) =\lVert \mathbf{x}_r- \mathbf{x}_p \rVert$. 

Our model predicts each multipole position $\mathbf{x}_p$, its emitting signal $s_p(t) = \mathcal{F}^{-1}[S_p(f)]$, as well as its spherical harmonic coefficients $b_{nm,p}(t)= \mathcal{F}^{-1}[B_{nm,p}(f)]$, for the given receiver position $\mathbf{x}_r$ and inverse Fourier transform $\mathcal{F}^{-1}$. The RIR is then synthesized in the frequency domain from Eq.~\eqref{eq:model_f} using a set of estimated parameters $\Theta = \{(\mathbf{x}_p, S_p(f), B_{nm,p}(f))\}_{p=1}^P$. The model is trained to minimize a loss function comparing the ground-truth and estimated RIRs (detailed loss functions are described in Section \ref{sec:implementation}). 

The NAMS architecture to estimate the parameter set $\Theta$ is depicted in Fig.~\ref{fig:placeholder}. The architecture consists of signal and directivity branches for estimating multipole position and signal pairs $\{\mathbf{x}_p, s_p(t)\}_{p=1}^P$ and directional patterns $\{D_p(f,\mathbf{x}_r)\}_{p=1}^P$, respectively. In the upper branch, a set of multipole positions $\{\mathbf{x}_p\}$ is declared as learnable network parameters, and only this information is utilized to synthesize the multipole signal $s_p(t)$. This process is to ensure that the multipole signals are independent of receiver positions. Specifically, the multipole position is encoded by a positional encoder and then fed into the signal estimation layer (MLP) to generate a short-length signal $s_p(t)$ in the time domain. The multipole index $p$ is assigned to the batch dimension, and the MLP is trained to generate source signals corresponding to their multipole positions.
In the lower branch, the model takes the receiver position $\mathbf{x}_r$ as input and subtracts it from multipole positions $\mathbf{x}_p$ to calculate the relative positions of multipoles required for deriving $\mathbf{\Omega}_{p}(\mathbf{x}_r)$.  
The relative multipole positions are also encoded by a positional encoder and concatenated with the encoded $\mathbf{x}_p$. From the encoded position vectors, the directivity estimation layer (MLP) synthesizes the spherical harmonic coefficients $\{b_{nm,p}(t)= \mathcal{F}^{-1}[B_{nm,p}(f)]\}$ in the time domain.  

The coefficients are then used to generate the directional patterns according to Eq.~\eqref{eq:dir_f}. We use real-valued spherical harmonics and constrain the total energy of $D_p(f,\mathbf{x}_r)$ across frequencies to stabilize the training process. In detail, for a vector $\mathbf{D}_p = \left[D_p(f_1,\mathbf{x}_r),~\cdots,~D_p(f_F,\mathbf{x}_r)\right]$ defined for discrete frequencies $f_1,\cdots, f_F$, we apply normalization $\tilde{\mathbf{D}}_p={\mathbf{D}_p}/{\left\lVert \mathbf{D}_p \right\rVert}$ and use it in place of $D_p(f,\mathbf{x}_r)$ in Eq.~\eqref{eq:model_f}.
The entire process described above is differentiable, so we can train both $\{\mathbf{x}_p\}_{p=1}^P$ and the MLPs via backpropagation. Moreover, the physical constraint on the Helmholtz equation is automatically satisfied, because multipoles constituting Eq.~\eqref{eq:model_f} are the solution of the Helmholtz equation.

\subsection{Pruning Neural Multipoles}
\label{pruning}
An excessive number of multipoles can lead to overfitting, significantly degrading computational efficiency. Therefore, it is essential to determine the optimal number of multipoles. To achieve this, we densely distribute the multipoles in the space at model initialization and incorporate pruning stages during training.
We regularly perform pruning every 20 epochs after the first 100 epochs of training. 
The principle of pruning is to remove multipoles with low $s_p(t)$ energy. To compute the energy of $s_p(t)$ for each multipole, we freeze the model and input a dummy receiver position. Since $s_p(t)$ depends solely on $\mathbf{x}_p$, the choice of dummy receiver position does not affect the result. Also, the total energy of $D_p(f,\mathbf{x}_r)$ is constrained to one, so we can determine the necessity of each multipole solely by the energy of $s_p(t)$.  
If the calculated energy is below 50\% of the global median, the corresponding $\mathbf{x}_p$ is removed from the model.
After each pruning step, the model re-optimizes $\{\mathbf{x}_p\}_{p=1}^{P'}$ and estimation layers using the reduced number ($P'$) of multipoles. 
This iterative alternating process allows for the development of a compact NAMS model with less possibility of overfitting.

\section{Experiment}
\label{sec:Experiment}

\subsection{Datasets}
\label{sec:dataset}
\begin{figure}[h]
    \centering
    \begin{adjustbox}{minipage=\linewidth,scale=0.95}
    \begin{subfigure}[b]{0.5\linewidth}
        \centering
        \includegraphics[width=\linewidth]{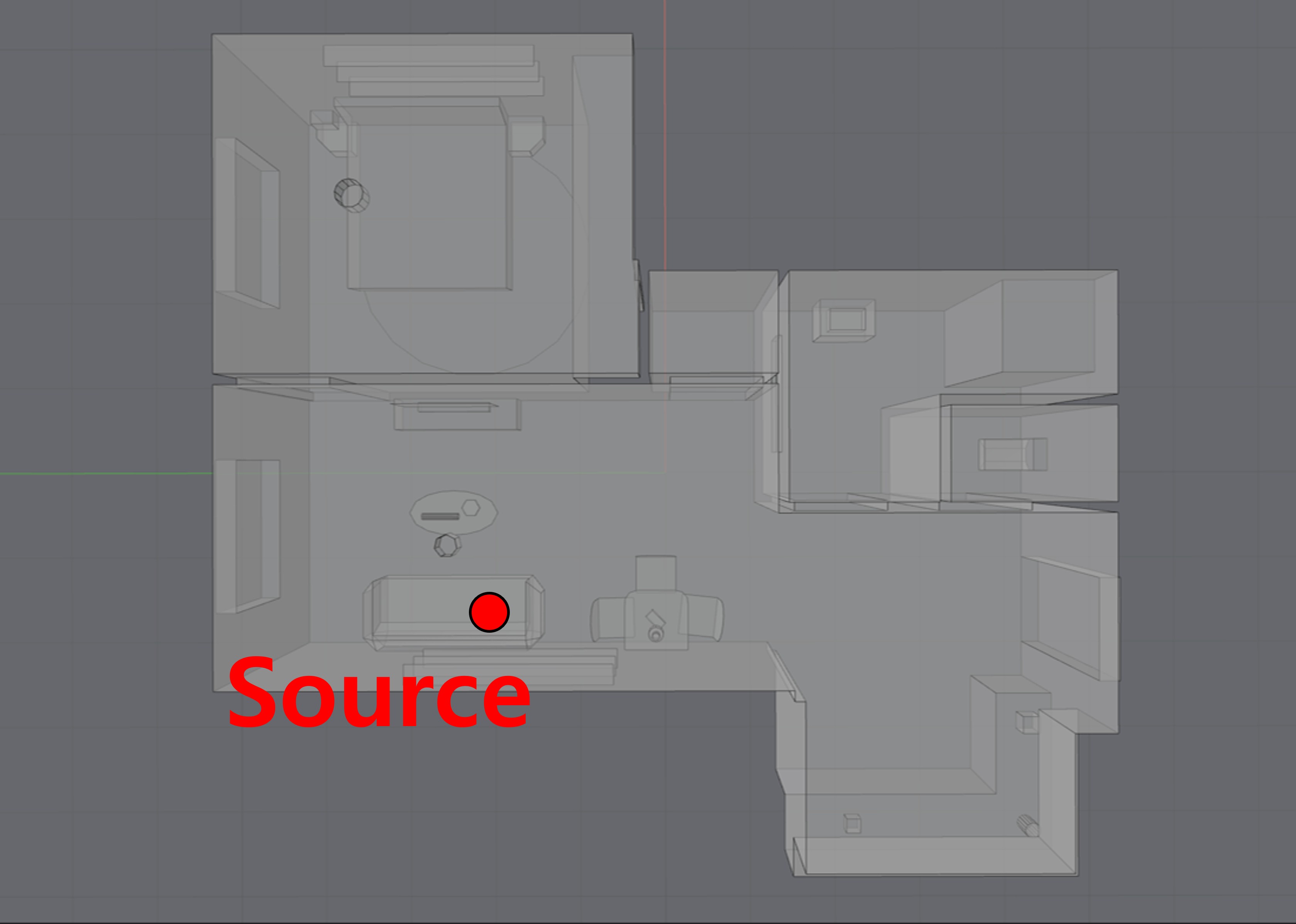}
        \label{fig:left} \vspace{-10 pt}
        \caption{Apartment 566}
    \end{subfigure}
    \hfill
    \begin{subfigure}[b]{0.5\linewidth}
        \centering
        \includegraphics[width=\linewidth]{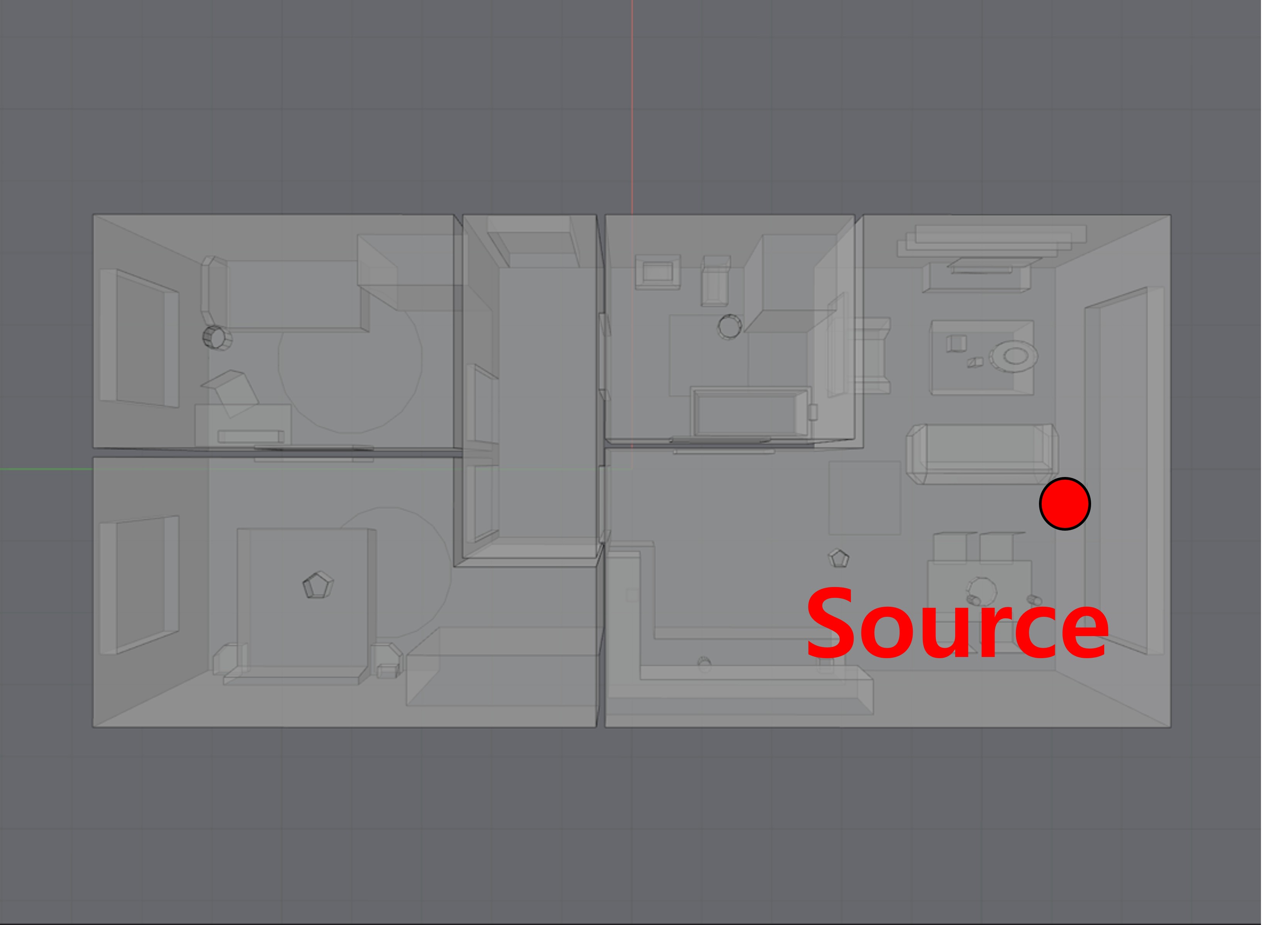}
        \label{fig:right} \vspace{-10 pt}
        \caption{Apartment 716}
    \end{subfigure}
    \end{adjustbox}
    \vspace{-6 pt}
    \caption{Room and source configurations used for the simulations.}
    \label{fig:treble_room}
    \vspace{-10 pt}
\end{figure}

Our model's performance was assessed using both real and synthetic datasets. For the real dataset, we utilized the MeshRIR dataset~\cite{koyama2021meshrir}. As for the synthetic dataset, we conducted simulations of two room environments employing the Treble simulator~\footnote{https://www.treble.tech}, which offers a hybrid simulation approach integrating wave-based and geometrical acoustics. Specifically, we chose two scenes: Apartment 566 and Apartment 716 from the Treble database. The scenes contain furniture such as sofas, carpets, and beds, with absorption and scattering coefficients taken from the predefined values in the Treble database. Apartments 566 and 716 have volumes of 105 m$^3$ and 183 m$^3$, respectively, and follow a Manhattan layout. The reverberation times (T60) of Apartments 566 and 716, averaged over all receiver positions, were 0.48 s and 1.80 s, respectively. We positioned a single omnidirectional source at a fixed location selected at random. RIRs were simulated for 1,000 randomly chosen receiver positions. Fig.~\ref{fig:treble_room} depicts the employed room layouts. Across all experiments, we randomly split the RIRs into training and testing sets with a ratio of 9:1. All RIRs were resampled to 24 kHz sampling rate and trimmed to 0.1 seconds.

\subsection{Implementation Details}
\label{sec:implementation}
During the initial phase, we densely splatted multipoles in space. The starting locations were arranged as points on a collection of spheres, each centered on the source, with radii incrementing from 1 m to 34 m at intervals of 1 m. On each sphere, 32 points were uniformly distributed using Fibonacci sampling and randomly rotated together. An additional multipole was located at the source position to render the direct sound, yielding a total of 1,089 initial positions. 
We applied sinusoidal positional embeddings with 10 sine–cosine frequencies to encode source and receiver positions, following~\cite{luo2022learning}. A 3-layer MLP with 512 hidden units per layer was used to predict $s_p(t)$, and the same architecture was employed to predict $b_{nm,p}(t)$. The output $s_p(t)$ has a duration of 3 ms, and $b_{nm,p}(t)$ includes spherical harmonic coefficients up to the 3rd order, yielding a 16-channel output with a duration of 3 ms.

For optimization, we adopted the loss function used in AVR~\cite{lan2024acoustic}, which consists of a weighted sum of spectral loss, amplitude loss, phase loss, time-domain loss, multi-resolution STFT loss~\cite{yamamoto2020parallel}, and energy decay loss~\cite{majumder2022few}, with weights of 1, 0.5, 0.5, 100, 1, and 5, respectively. We used the Adam optimizer with a cosine scheduler, decaying the learning rate from $10^{-3}$ to $10^{-4}$. All experiments were conducted for 300 epochs, and the best model was updated at the epoch whenever the test loss was minimized. All experiments were run on a single RTX A6000 GPU.


\begin{table*}[t]
\centering
\scriptsize
\setlength{\tabcolsep}{8.65pt}
\renewcommand{\arraystretch}{1.4}
\begin{tabular}{l|ccccc|ccccc|ccccc}
\hline
\multirow{2}{*}{\textbf{Method}} 
& \multicolumn{5}{c|}{\textbf{MeshRIR}} 
& \multicolumn{5}{c|}{\textbf{Apartment 566}}
& \multicolumn{5}{c}{\textbf{Apartment 716}} \\
\cline{2-16}
& Amp. & Env. & T60 & C50 & EDT  
& Amp. & Env. & T60 & C50 & EDT 
& Amp. & Env. & T60 & C50 & EDT \\
\hline
NAF\cite{luo2022learning} & 0.57 & 1.98 & 3.5 & 0.88 & 31.0 
        & 0.77 & 5.05 & 15.2 & 8.66 & 22.0 
        & 0.76 & 6.56 & 21.4 & 7.35 & 18.8 \\
AVR\cite{lan2024acoustic} & 0.28 & 1.44 & 2.9 & 0.66 & 19.4 
        & 0.46 & \textbf{4.20} & 5.0 & 1.20 & 29.8 
        & 0.54 & \textbf{5.79} & 9.0 & 2.46 & 24.2 \\
\rowcolor{red!15} NAMS & \textbf{0.11} & \textbf{1.21} & \textbf{2.0} & \textbf{0.34} & \textbf{9.8} 
         & \textbf{0.22} & 4.46 & \textbf{3.4} & \textbf{0.48} & \textbf{12.0} 
         & \textbf{0.30} & 6.70 & \textbf{6.3} & \textbf{0.82} & \textbf{13.7} \\
\hline
\end{tabular}
\caption{\textbf{RIR estimation performance comparison with existing models}. The evaluation metrics include the amplitude error (Amp.), envelope error (Env., \%), relative T60 error (\%), C50 error (dB), and EDT error (ms). }
\label{tab:modelcomparison}
\end{table*}


\begin{table*}[t]
\centering
\scriptsize
\setlength{\tabcolsep}{4pt}
\renewcommand{\arraystretch}{1.2}
\begin{tabular}{l|ccccc|cc|ccccc|cc|ccccc|cc}
\hline
\multirow{2}{*}{\textbf{Method}} 
& \multicolumn{7}{c|}{\textbf{MeshRIR}} 
& \multicolumn{7}{c|}{\textbf{Apartment 566}}
& \multicolumn{7}{c}{\textbf{Apartment 716}} \\
\cline{2-22}
& Amp. & Env. & T60 & C50 & EDT & \#pts & $\mathrm{T}_\mathrm{Inf}$
& Amp. & Env. & T60 & C50 & EDT & \#pts & $\mathrm{T}_\mathrm{Inf}$
& Amp. & Env. & T60 & C50 & EDT & \#pts & $\mathrm{T}_\mathrm{Inf}$ \\
\hline
mono. sparse & 0.19 & 1.36 & 2.3 & 0.39 & 12.4 & 273 & \textbf{1.7}
        & 0.26 & 4.81 & 4.0 & 0.60 & 16.0 & 307 & \textbf{1.7}
        & 0.59 & 10.25 & 7.8 & 0.82 & 16.1 & 273 & \textbf{1.7}  \\
\rowcolor{gray!15} multi. sparse & 0.15 & 1.30 & 2.1 & 0.37 & 11.5 & 273 & 2.1
        & \textbf{0.22} & 4.50 & 3.6 & \textbf{0.43} & \textbf{11.5} & 307 & 2.2
        & 0.40 & 8.20 & \textbf{5.9} & \textbf{0.76} & \textbf{13.5} & 273 & 2.1  \\
\hline
mono. dense & 0.15 & 1.27 & 2.2 & 0.37 & 10.8 & 1089 & 2.0
        & 0.28 & 4.89 & 4.0 & 0.69 & 17.0 & 1089 & 2.0
        & 0.50 & 9.33 & 7.2 & 0.99 & 16.4 & 1089 & 2.0  \\

\rowcolor{gray!15} multi. dense & 0.12 & \textbf{1.20} & \textbf{2.0} & \textbf{0.33} & 10.2 & 1089 & 4.5
        & \textbf{0.22} & \textbf{4.39} & 3.9 & 0.50 & 13.3 & 1089 & 4.6
        & 0.42 & 8.47 & 6.3 & 0.87 & 13.6 & 1089 & 4.6  \\

\hline

\rowcolor{red!15} \makecell{multi. dense \\ \textit{w/ pruning}} 
& \textbf{0.11} & 1.21 & \textbf{2.0} & 0.34 & \textbf{9.8} & \textbf{225} & 2.2 
& \textbf{0.22} & 4.46 & \textbf{3.4} & 0.48 & 12.0 & \textbf{276} & 2.2 
& \textbf{0.30} & \textbf{6.70} & 6.3 & 0.82 & 13.7 & \textbf{240} & 2.1 \\
        
\hline
\end{tabular}
\caption{\textbf{Effectiveness of multipole and pruning.} Performance comparison of monopole and multipole splatting. 'mono.' and 'multi.' represent monopole and multipole settings, respectively, whereas 'sparse' and 'dense' denote the small and large number of poles at initialization. 
}

\label{tab:ablation1}
\end{table*}

\begin{table}[t!]
\centering
\scriptsize
\setlength{\tabcolsep}{5.pt}
\renewcommand{\arraystretch}{1.4}
\begin{tabular}{l|ccc|ccc}
\hline
\multirow{2}{*}{\textbf{Method}} 
& \multicolumn{3}{c|}{\textbf{MeshRIR}} 
& \multicolumn{3}{c}{\textbf{Apartment 566 \& 716}}\\
\cline{2-7}
  & Param. & \#pts & $\mathrm{T}_\mathrm{Inf}$  & Param. & \#pts & $\mathrm{T}_\mathrm{Inf}$\\
\hline
NAF\cite{luo2022learning} & 2.7M & - & \textbf{1.9} & 2.7M &  - & \textbf{1.95} \\
AVR\cite{lan2024acoustic} & 57.2M (2.0M) & 205k & 62.5 & 57.2M (2.0M) & 205k & 61.9   \\
\rowcolor{red!15}NAMS & \textbf{1.8M} & \textbf{225} & 2.2 & \textbf{1.8M} & \textbf{258} & 2.15\\
\hline
\end{tabular}
\caption{\textbf{Space and time complexity.} Comparison of the model parameter size (Param.), number of sample points or multipoles (\#pts), and inference time ($\mathrm{T}_\mathrm{Inf}$, ms).}
\label{tab:modelspace}
\end{table}

\begin{figure}[h]
    \centering
    \includegraphics[width=1\columnwidth]{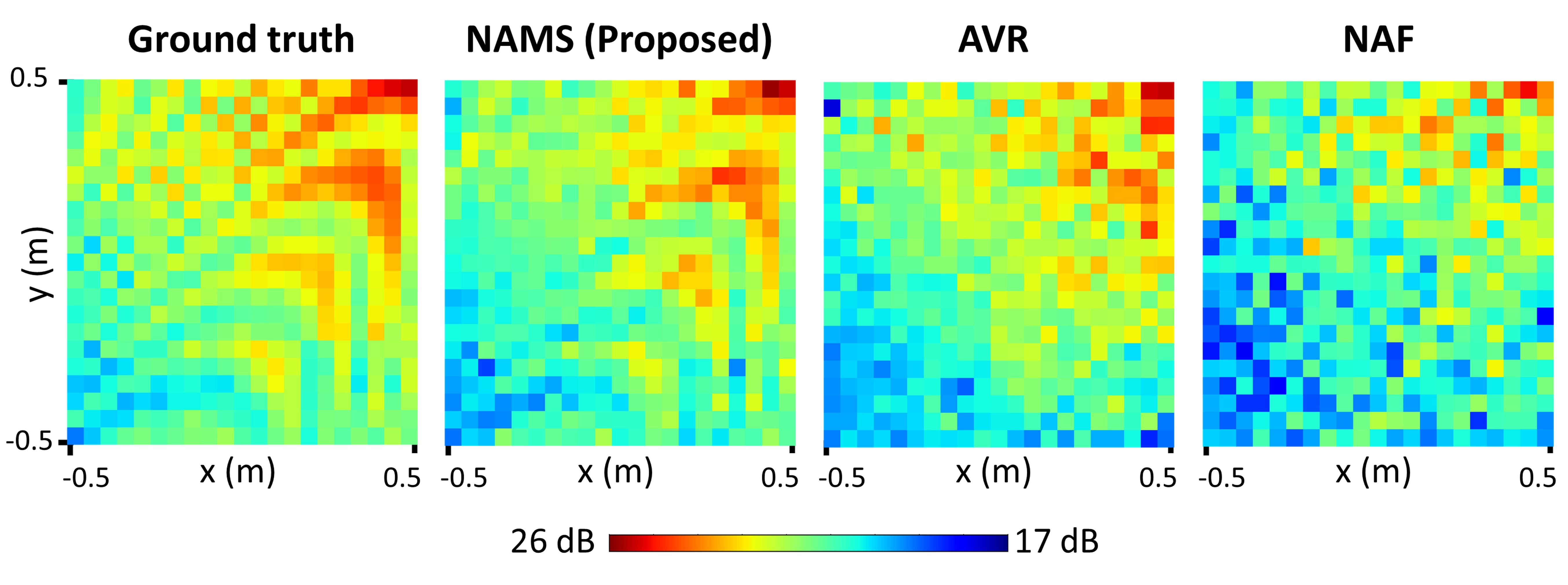}
    \vspace{-15 pt}
    \caption{\textbf{Comparison of spatial magnitude distributions.} Overall magnitude distribution averaged over the 1/3 octave band centered at 4 kHz (Scene from MeshRIR).}
    \label{fig:spatial_magnitude}
\end{figure}


\begin{figure}[h]
    \centering
    \includegraphics[width=1\linewidth]{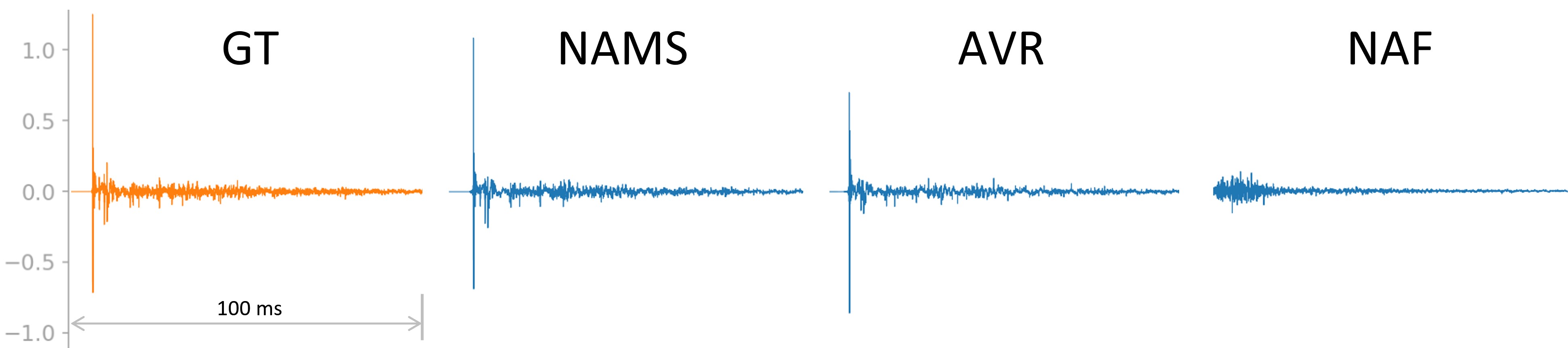}
    \caption{\textbf{Ground-truth vs. generated RIRs.} Samples with comparable amplitude errors to Table~\ref{tab:modelcomparison} are selected from MeshRIR.}
    \label{fig:ir_comparison}
\end{figure}


\begin{figure}[h]
    \centering
    \begin{subfigure}{0.32\linewidth}
        \centering
        \includegraphics[width=\linewidth]{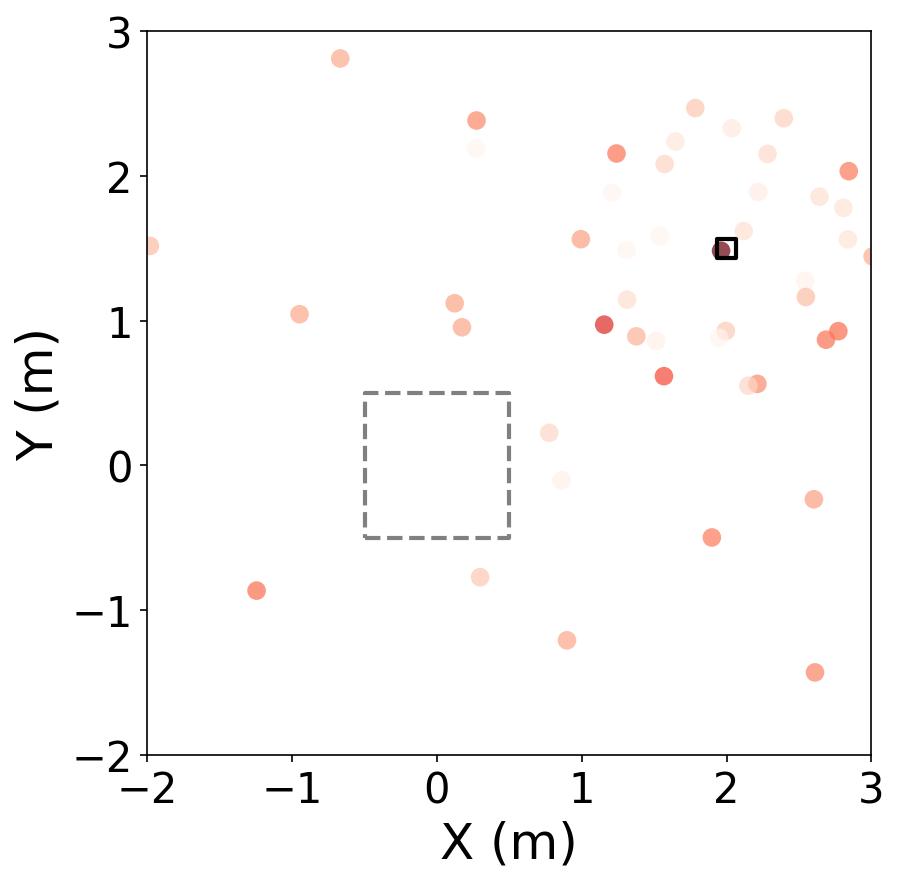}
        \caption{Dense w/o pruning}
        \label{fig:pruning_a}
    \end{subfigure}
    \hfill
    \begin{subfigure}{0.32\linewidth}
        \centering
        \includegraphics[width=\linewidth]{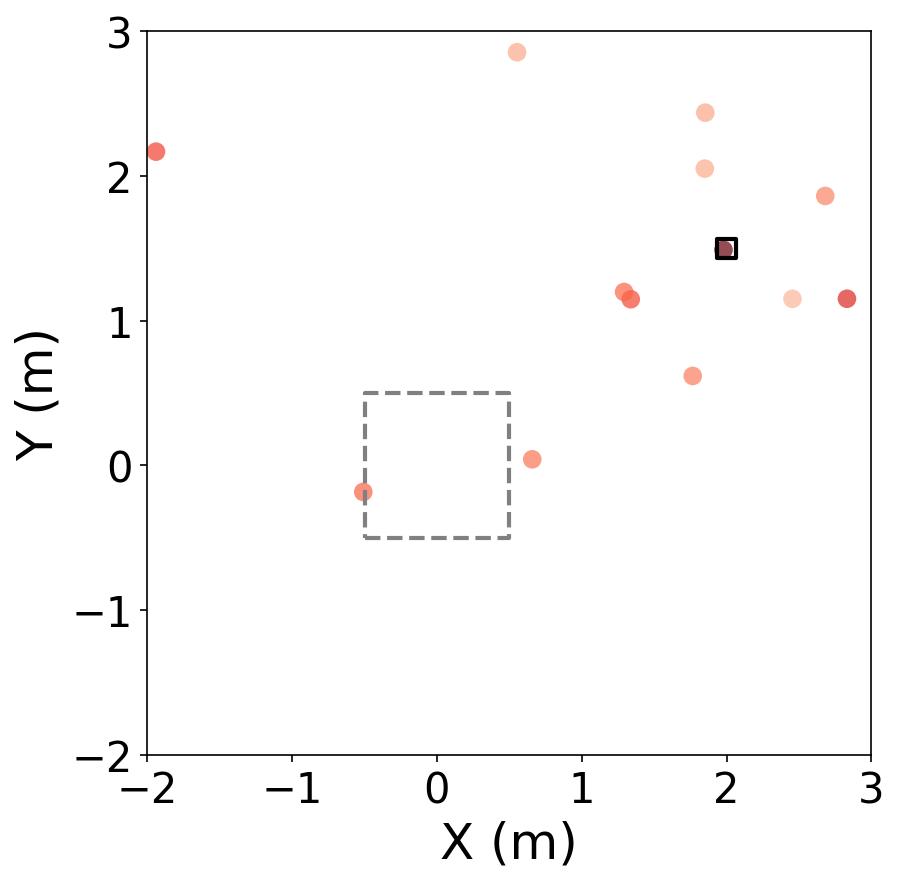}
        \caption{Sparse w/o pruning}
        \label{fig:pruning_b}
    \end{subfigure}
    \hfill
    \begin{subfigure}{0.32\linewidth}
        \centering
        \includegraphics[width=\linewidth]{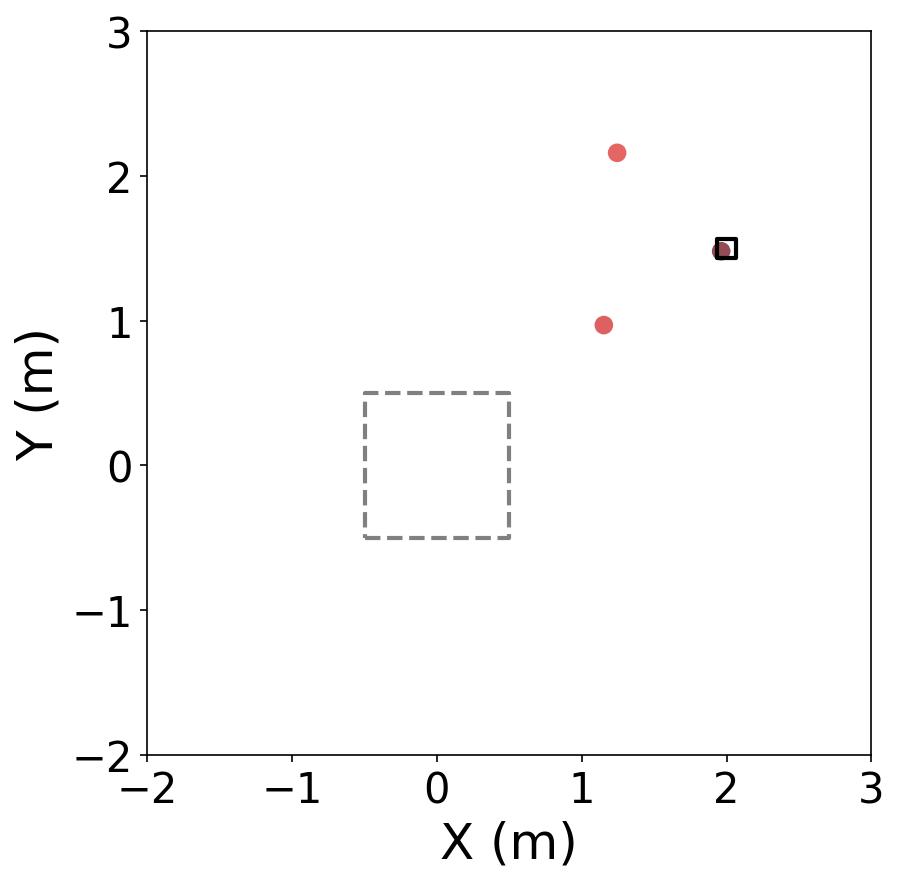}
        \caption{Dense w/ pruning}
        \label{fig:pruning_c}
    \end{subfigure}
    \vspace{-5pt}
    \caption{\textbf{Multipole positions and signal energies with and without pruning.} Circles indicate the multipoles inside the room, with darker red indicating higher energy. The black square denotes the source, and the gray dashed square indicates the measurement region.}
    \label{fig:pruning}
\end{figure}


\section{Results}
\label{sec:results}

\subsection{Performance Comparison}
\label{sec: performance comparison}

We compared our model with NAF~\cite{luo2022learning} and AVR~\cite{lan2024acoustic} that estimate RIRs only from the given source and receiver positions. For a fair comparison, all models were trained for the same source position in the real and synthetic datasets. 
The point-neuron model~\cite{bi2024point} was excluded from the comparison because its code is not publicly available. Instead, in our ablation study, we evaluated performance differences between monopole and multipole configurations.

The model performances were assessed using the following metrics: amplitude error, envelope error, reverberation time (T60) error, clarity (C50) error, and early decay time (EDT) error~\cite{lan2024acoustic}. Amplitude error is relative error with no unit; envelope error is presented as a relative percentage error, and the units for T60, C50, and EDT errors are \%, dB, and milliseconds, respectively.

We present the experimental results in Table~\ref{tab:modelcomparison}. On the MeshRIR dataset, NAMS outperforms existing methods on all metrics. Similar trends can be seen on the Apartment 566 and 716 datasets, except for the envelope error. 
To visualize the differences in the estimated sound field, we present the spatial magnitude distribution of sound fields estimated from the MeshRIR dataset. The results show a clear difference of NAMS compared to baseline models with less spatial jitter. In terms of RIR waveforms shown in Fig.~\ref{fig:ir_comparison}, NAMS captures the direct sound and early reflection peaks more accurately than AVR and NAF.

In Table~\ref{tab:modelspace}, we also compare the space and time complexity of models. NAMS also shows a fast inference time ($\mathrm{T}_\mathrm{Inf}$) of 2.1--2.2 ms, comparable to NAF (1.9--2.0 ms). NAMS achieves this speed using only a few hundred multipoles, whereas AVR requires sampling 204,800 points. In addition, our model has an advantage in terms of parameter size. AVR~\cite{lan2024acoustic} has 57.2 million parameters, whereas NAMS and NAF~\cite{luo2022learning} have 1.8 million and 2.7 million parameters, respectively. However, the large parameter count of AVR~\cite{lan2024acoustic} is mainly due to its use of hash grid encoding~\cite{muller2022instant}. Excluding the hash grid encoding, AVR has 2.0 million parameters, slightly higher than NAMS.

\subsection{Ablation studies and discussions}
We conducted ablation studies to investigate the efficacy of multipole splatting by comparing the monopole and multipole models. The monopole model is implemented by setting the maximum order ($N$) of spherical harmonics to zero. We also considered two different pole initializations for each configuration. The dense initialization denotes the initialization described in Section \ref{sec:implementation}, while the sparse initialization indicates the case where the number of points on each sphere is reduced such that the total number of poles is similar to that obtained after pruning the dense initialization model. The results are shown in Table \ref{tab:ablation1}. 

For the same number of poles, multipole models consistently outperform monopole models across all datasets. Despite this, multipole models incur longer inference time due to their larger number of trainable parameters. Nevertheless, in both the Apartment 566 and 716 datasets, the sparse multipole model outperforms all monopole models, including the dense initialization model. This result indicates that multipoles offer a more expressive representation in complex acoustic environments. 
Conversely, on the MeshRIR dataset, both the multipole model with sparse initialization and the monopole model with dense initialization perform similarly, suggesting that monopoles might suffice for modeling uncomplicated, unobstructed rooms.

As indicated in Table~\ref{tab:ablation1}, pruning decreases the multipoles to 225, 276, and 240 in the MeshRIR, Apartment 566, and Apartment 716 datasets, respectively, while still outperforming the dense monopole model with 1,089 poles. This highlights that the multipole model more effectively represents sound fields using only 20--22\% of poles. The pruned model also resulted in over twice the inference speed compared to the non-pruned dense multipole initialization for each dataset. Moreover, the compact model aids in interpreting the physical structure of a sound field. To illustrate, we present the multipole distributions optimized for the MeshRIR dataset under three scenarios (Fig.~\ref{fig:pruning}): dense and sparse splatting without pruning, and dense splatting with pruning. In the pruned dense setup, three high-energy multipoles are placed near the source, whereas both dense and sparse setups without pruning spread many low-energy multipoles in the space. Thus, the pruned model offers a more interpretable, structured portrayal of a sound field using a limited set of multipoles with adaptable directivities and signal emission. Additional demos are available at \url{https://bgw6287.github.io/nams-demo/}.

\section{Conclusion}
We introduced NAMS to synthesize room impulse responses via multipoles, using pruning to auto-select the optimal quantity. This approach outperforms existing methods while ensuring fast inference by efficiently representing sound fields with refined multipole directivities and placements. We demonstrated that multipoles offer a richer representation than monopoles in complex acoustic environments. Additionally, starting with a dense multipole set and optimizing through pruning achieves better results than manual initialization. This evidence shows NAMS efficiently represents RIRs. The next step for practical RIR estimation will be designing a generalized model for various source positions using fewer RIRs.



\vfill\pagebreak



{\fontsize{9pt}{10.7pt}\selectfont
\section{Acknowledgements}
This work was supported by the National Research Foundation of Korea (NRF) grant (No. RS-2024-00337945) and the BK21 FOUR program through the NRF grant funded by the Ministry of Education of the Korean government (MOE).

\bibliographystyle{IEEEbib}
\bibliography{strings,refs}
}

\end{document}